\documentclass[floatfix,twocolumn,showpacs,preprintnumbers,amsmath,amssymb,pra,superscriptaddress,longbibliography]{revtex4-1}
\usepackage{color}
\usepackage[usenames,dvipsnames,svgnames,table]{xcolor}
\usepackage[colorlinks=true,linkcolor=blue,urlcolor=blue,citecolor=blue]{hyperref}
\usepackage{mathtools}
\usepackage{graphicx}
\usepackage{dcolumn}
\usepackage{array}
\usepackage{lipsum}
\usepackage{bm}
\usepackage{subfigure}
\usepackage{amssymb}
\usepackage{multirow}
\usepackage{tabularx}
\usepackage{amsmath}
\usepackage{braket}
\usepackage{csquotes}
\graphicspath{{plots/}}
 \usepackage{lipsum}
\usepackage{mathrsfs}
\usepackage{MnSymbol}
	
\newcommand{\beq}{\begin{equation}}
\newcommand{\eeq}{\end{equation}}
\newcommand{\bea}{\begin{eqnarray}}
\newcommand{\eea}{\end{eqnarray}}

\begin{document}
\title{\emph{Ab initio} density functional theory approach to warm dense hydrogen:\\ from density response to electronic correlations}

\author{Zhandos~A.~Moldabekov}
\email{z.moldabekov@hzdr.de}
\affiliation{Center for Advanced Systems Understanding (CASUS), D-02826 G\"orlitz, Germany}
\affiliation{Helmholtz-Zentrum Dresden-Rossendorf (HZDR), D-01328 Dresden, Germany}

\author{Xuecheng Shao}
\affiliation{
Key Laboratory of Material Simulation Methods $\&$ Software of Ministry of Education, College of Physics, Jilin University, Changchun 130012, PR China}
\affiliation{State Key Laboratory of High Pressure and Superhard Materials, Jilin University, Changchun 130012, PR China }

\author{Hannah M.~Bellenbaum}
\affiliation{Center for Advanced Systems Understanding (CASUS), D-02826 G\"orlitz, Germany}
\affiliation{Helmholtz-Zentrum Dresden-Rossendorf (HZDR), D-01328 Dresden, Germany}
\affiliation{Institut f\"ur Physik, Universit\"at Rostock, D-18057 Rostock, Germany}

\author{Cheng Ma}
\affiliation{
Key Laboratory of Material Simulation Methods $\&$ Software of Ministry of Education, College of Physics, Jilin University, Changchun 130012, PR China}
\affiliation{State Key Laboratory of High Pressure and Superhard Materials, Jilin University, Changchun 130012, PR China }

\author{Wenhui Mi}
\affiliation{
Key Laboratory of Material Simulation Methods $\&$ Software of Ministry of Education, College of Physics, Jilin University, Changchun 130012, PR China}
\affiliation{International Center of Future Science, Jilin University, Changchun 130012, PR China }
\affiliation{State Key Laboratory of High Pressure and Superhard Materials, Jilin University, Changchun 130012, PR China }

\author{Sebastian Schwalbe}
\affiliation{Center for Advanced Systems Understanding (CASUS), D-02826 G\"orlitz, Germany}
\affiliation{Helmholtz-Zentrum Dresden-Rossendorf (HZDR), D-01328 Dresden, Germany}

\author{Jan Vorberger}
\affiliation{Helmholtz-Zentrum Dresden-Rossendorf (HZDR), D-01328 Dresden, Germany}

\author{Tobias Dornheim}
\email{t.dornheim@hzdr.de}
\affiliation{Center for Advanced Systems Understanding (CASUS), D-02826 G\"orlitz, Germany}
\affiliation{Helmholtz-Zentrum Dresden-Rossendorf (HZDR), D-01328 Dresden, Germany}

\begin{abstract}
Understanding the properties of warm dense hydrogen is of key importance for the modeling of compact astrophysical objects and to understand and further optimize inertial confinement fusion (ICF) applications. The work horse of warm dense matter theory is given by thermal density functional theory (DFT), which, however, suffers from two limitations: (i) its accuracy can depend on the utilized exchange--correlation (XC) functional, which has to be approximated and (ii) it is generally limited to single-electron properties such as the density distribution. Here, we present a new ansatz combining time-dependent DFT results for the dynamic structure factor $S_{ee}(\mathbf{q},\omega)$ with static DFT results for the density response. This allows us to estimate the electron--electron static structure factor $S_{ee}(\mathbf{q})$ of warm dense hydrogen with high accuracy over a broad range of densities and temperatures. In addition to its value for the study of warm dense matter, our work opens up new avenues for the future study of electronic correlations exclusively within the framework of DFT for a host of applications.
\end{abstract}
\maketitle

Over the last decade, there has been a surge of interest in the properties of matter at extreme densities (Wigner-Seitz radius $r_s=d/a_\textnormal{B}\sim1$), temperatures ($\Theta=k_\textnormal{B}T/E_\textnormal{F}\sim1$, with $E_\textnormal{F}$ the electronic Fermi temperature~\cite{quantum_theory}) and pressures ($P\sim1-10^4\,$MBar)~\cite{vorberger2025roadmapwarmdensematter}. Such warm dense matter (WDM) conditions abound in a variety of compact astrophysical objects~\cite{wdm_book,fortov_review,drake2018high} such as giant planet interiors~\cite{Benuzzi_Mounaix_2014,Guillot2018}. The high-pressure moderate-temperature regime is of great interest for material synthesis and discovery, with the demonstration of the laser-driven formation of nanodiamonds by Kraus \emph{et al.}~\cite{Kraus2016,Kraus2017} being a case in point. A particularly intriguing technological application is given by inertial confinement fusion (ICF)~\cite{Betti2016,Hurricane_RevModPhys_2023}, where both the hydrogen fusion fuel and the ablator material have to traverse the WDM regime in a controlled way~\cite{hu_ICF}. Indeed, there have been a number of very recent spectacular experimental achievements~\cite{AbuShawareb_PRL_2024,Zylstra2022,Williams2024}, but the systematic optimization of fusion yields that is required to make ICF an economical source of clean energy~\cite{Batani_roadmap} will require integrated modeling with real predictive capability.

From a theoretical perspective, WDM is characterized by the complex interplay of effects such as Coulomb coupling, quantum degeneracy, thermal excitation, and partial ionization~\cite{new_POP,Bonitz_POP_2024,wdm_book}, which have to be taken into account holistically; this is a formidable challenge even for state-of-the-art computational methods. Thermal density functional theory (DFT)~\cite{Mermin_DFT_1965} has emerged as the workhorse of modern WDM theory, but it has two main limitations: (i) its accuracy decisively depends on the electronic exchange--correlation (XC) functional, which has to be approximated and supplied as an external input; (ii) it is, by construction, limited to single-electron observables such as the density distribution $n_e(\mathbf{r})$. There is a broad consensus on the importance of (i) in the community, which has resulted in a number of new XC-functionals that take into account temperature effects~\cite{ksdt,groth_prl,Karasiev_PRL_2018,Karasiev_PRB_2022,kozlowski2023generalized,Sjostrom_PRB_2014} and in first attempts at benchmarking different functionals at WDM conditions~\cite{karasiev_importance,kushal,moldabekov2024density,Moldabekov_JCTC_2024,Moldabekov_JCP_2023,Moldabekov_JCTC_2023}.
Here, we address point (ii) by introducing a new ansatz combining linear-response time-dependent DFT (TDDFT) results for the (inelastic) electronic dynamic structure factor $S_{ee}(\mathbf{q},\omega)$~\cite{Moldabekov_MRE_2025} with the static linear density response $\chi_{ee}(\mathbf{q})$ computed from the direct perturbation method~\cite{Moldabekov_JCTC_2022,Moldabekov_JCTC_2023,moldabekov2024density,Dornheim_review,moroni}.
This allows us to estimate the electron--electron static structure factor $S_{ee}(\mathbf{q})$---i.e., the Fourier transform of the pair correlation function $g_{ee}(\mathbf{r})$---without the need for any external input in addition to the usual XC-functional.
As a practical example, we consider warm dense hydrogen, which is of key importance for astrophysics and ICF applications alike~\cite{Bonitz_POP_2024}, and find excellent agreement with highly accurate path integral Monte Carlo (PIMC) reference data~\cite{Dornheim_MRE_2024,Dornheim_JCP_2024,bellenbaum2025estimatingionizationstatescontinuum} where they are available. In addition, we present new results for lower temperatures, where PIMC simulations break down due to the well-known fermion sign problem~\cite{dornheim_sign_problem,troyer}.

We expect our work to be of considerable interest for a gamut of future applications. First and foremost, we open up the way to study electronic correlations consistently within DFT on a true ab-initio level. This is directly relevant for the interpretation of experiments with WDM 
where $S_{ee}(\mathbf{q})$ is probed both with x-ray diffraction and x-ray Thomson scattering diagnostics~\cite{siegfried_review,Dornheim_SciRep_2024,Dornheim_review,sheffield2010plasma,Dornheim_Science_2024}.
In addition, our work might give new insights into the construction of advanced XC-functionals based on the combination of the adiabatic connection formula with the fluctuation--dissipation theorem~\cite{Fuchs_PRB_2002,pribram}. 
Finally, our set-up has great potential to benchmark and constrain dynamic XC-kernels~\cite{Ruzsinszky_PRB_2020,Panholzer_PRL_2018,Dornheim_PRB_2024,koskelo2023shortrange} and to guide novel developments in the field of TDDFT.

\begin{figure}
\centering
\includegraphics[width=0.47\textwidth]{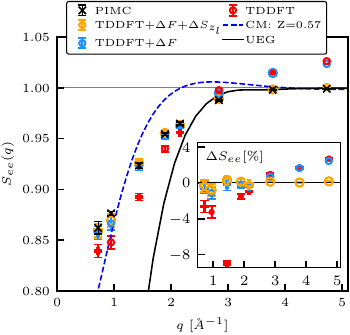}
\caption{\label{fig3} Electronic static structure factor $S_{ee}(\mathbf{q})$ of hydrogen at $\rho=0.08\,$g/cc [$r_s=3.23$] and $T=4.8\,$eV [$\Theta=1$] for $N=32$ (circles) and $N=14$ (plusses) atoms. Black crosses: quasi-exact PIMC reference data~\cite{Dornheim_JCP_2024}; red symbols: direct TDDFT; green symbols: TDDFT + $\chi_A$-correction [Eq.~(\ref{eq:correction})]; yellow symbols: TDDFT + $\chi_A(\mathbf{q})$ and $\Delta\chi_l$ [Eq.~(\ref{eq:quantum_correction})] corrections; dashed blue: chemical model~\cite{bellenbaum2025estimatingionizationstatescontinuum}, see Supplemental Material~\cite{supplement}. The inset shows the relative deviation of TDDFT data sets with respect to PIMC.
}
\end{figure}

\begin{figure}
\centering
\includegraphics[width=0.437\textwidth]{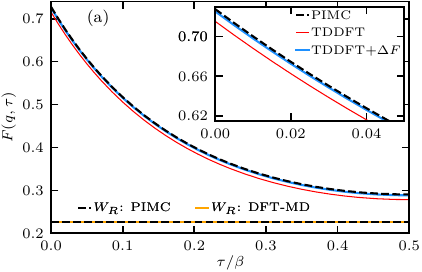}
\includegraphics[width=0.437\textwidth]{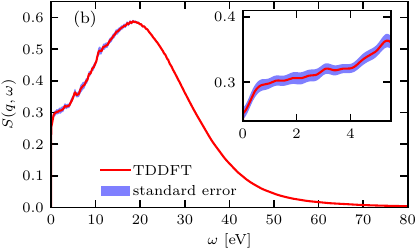}
\caption{\label{fig1}
(a) Inelastic ITCF $F_\textnormal{inel}(\mathbf{q},\tau)$ (top curves) and elastic ITCF $F_\textnormal{el}(\mathbf{q},\tau)=W_R(\mathbf{q})$ (horizontal lines) computed from PIMC (dashed black), direct TDDFT (solid red) and TDDFT + $\chi_A$ correction [Eq.~(\ref{eq:correction})] (solid green) for $T=4.8\,$eV, $\rho=0.08\,$g/cc and $q=1.892$\AA$^{-1}$. (b) Corresponding dynamic structure factor $S_\textnormal{inel}(\mathbf{q},\omega)$, with the inset showing a magnified segment around $\omega\to0$.
}
\end{figure}

\textbf{Results.} It has become a common practice to consider the imaginary-time correlation function (ITCF)~\cite{Dornheim_MRE_2023,Dornheim_review,Schoerner_PRE_2023,Dornheim_T_2022,Dornheim_T_follow_up,shi2025firstprinciplesanalysiswarmdense,Dornheim_PTR_2023}
\begin{eqnarray}\label{eq:Laplace}
    F_{ee}(\mathbf{q},\tau) = \int_{-\infty}^\infty \textnormal{d}\omega\ S_{ee}(\mathbf{q},\omega)\ e^{-\tau\hbar\omega}\ ,
\end{eqnarray}
which is given by the two-sided Laplace transform of the dynamic structure factor $S_{ee}(\mathbf{q},\omega)$; the sought after static structure factor is given by $S_{ee}(\mathbf{q})=F_{ee}(\mathbf{q},0)$.
Next, we decompose the full dynamic structure factor into an elastic and inelastic part,
\begin{eqnarray}\label{eq:define_WR}
    S_{ee}(\mathbf{q},\omega) = \underbrace{W_R(\mathbf{q}) \delta(\omega)}_{S_\textnormal{el}(\mathbf{q},\omega)} + S_\textnormal{inel}(\mathbf{q},\omega)\ ,
\end{eqnarray}
with $W_R(\mathbf{q})$ being the Rayleigh weight~\cite{Vorberger_PRE_2015,dornheim2024modelfreerayleighweightxray} characterizing the electronic localization around the nuclei; note that Eq.~(\ref{eq:define_WR}) exclusively relies on the higher nuclei mass compared to the electrons, and does not presuppose an additional decomposition into bound and free electrons~\cite{Chihara_1987,Gregori_PRE_2003}. The Rayleigh weight constitutes a standard observable in DFT molecular dynamics calculations~\cite{Vorberger_PRE_2015,dornheim2024modelfreerayleighweightxray}, and $S_\textnormal{inel}(\mathbf{q},\omega)$ is computed within the framework of linear-response TDDFT using the efficient Liouville-Lanczos method~\cite{Moldabekov_MRE_2025} with the PBE XC-functional~\cite{Perdew_PRL_1996} and the corresponding adiabatic XC-kernel; see the Supplemental Material~\cite{supplement} for additional details.

Evaluating Eq.~(\ref{eq:Laplace}) for $\tau=0$ gives us the direct TDDFT results for $S_{ee}(\mathbf{q})$, which are shown as the red circles in Fig.~\ref{fig3} for solid-density hydrogen ($\rho=0.08\,$g/cc) at the electronic Fermi temperature of $T=4.8\,$eV. These conditions can be realized in experiments with hydrogen jets~\cite{Zastrau,Fletcher_Frontiers_2022} and are also important for the initial phase of ICF experiments with cryogenic fuel~\cite{AbuShawareb_PRL_2024}. In addition, an ionization degree of $Z\sim0.5$ has been reported in the literature~\cite{bellenbaum2025estimatingionizationstatescontinuum,Militzer_PRE_2001,Bohme_PRL_2022}, making these conditions particularly interesting.
The black crosses depict exact PIMC reference data~\cite{Dornheim_MRE_2024,Dornheim_JCP_2024}, which provide an unassailable baseline for other theories.
For completeness, we have also included corresponding results for a uniform electron gas (UEG)~\cite{review,Dornheim_PRL_2020_ESA} at the same conditions (solid black). For $q\lesssim3q_\textnormal{F}$, the UEG static structure factor exhibits a sharp drop due to the perfect screening in the one-component plasma~\cite{kugler_bounds}; in contrast, the PIMC results for hydrogen attain a finite value in the limit of $q\to0$ in accordance with the compressibility sum rule~\cite{Vorberger_PRE_2015}.

Let us next consider the red crosses and plusses, which show direct TDDFT results for $N=32$ and $N=14$ hydrogen atoms. We find no systematic finite-size effects, which is a general feature of $q$-resolved properties~\cite{Chiesa_PRL_2006,dornheim_prl,Drummond_PRB_2008,Holzmann_PRB_2016,Dornheim_JCP_2021} at these conditions.
Overall, the direct TDDFT results capture the correct qualitative trend of $S_{ee}(\mathbf{q})$ for $q\lesssim3q_\textnormal{F}$, which is shaped by the electronic localization around the ions.
At the same time, we find significant deviations over the entire range of wavenumbers (deviations with respect to PIMC are shown in the inset of Fig.~\ref{fig3}), which are comparable in magnitude to a more simple chemical Chihara model~\cite{bellenbaum2025estimatingionizationstatescontinuum}, see the dashed blue line.

To understand the origin of the limited accuracy of the direct TDDFT results, we follow
Ref.~\cite{Moldabekov_MRE_2025} and transform our TDDFT results for $S_\textnormal{inel}(\mathbf{q},\omega)$ to the $\tau$-domain via Eq.~(\ref{eq:Laplace}). A corresponding comparison against exact PIMC reference data~\cite{Dornheim_MRE_2024,Dornheim_JCP_2024,bellenbaum2025estimatingionizationstatescontinuum} for $F_\textnormal{inel}(\mathbf{q},\tau)=F_{ee}(\mathbf{q},\tau)-W_R(\mathbf{q})$ and $F_\textnormal{el}(\mathbf{q},\tau)=W_\textnormal{R}(\mathbf{q})$ is shown in Fig.~\ref{fig1} (a) for $q=1.89$\AA$^{-1}$. While PIMC and DFT results for the latter are basically indistinguishable, we find excellent agreement over the entire $\tau$-range for $F_\textnormal{inel}(\mathbf{q},\tau)$, except for a constant shift; this must be connected with the $\omega\to0$ limit of $S_\textnormal{inel}(\mathbf{q},\omega)$~\cite{Dornheim_MRE_2023,bellenbaum2025estimatingionizationstatescontinuum}, as any effect at finite $\omega$ would affect the $\tau$-dependence of $F(\mathbf{q},\tau)$. 
The corresponding TDDFT results for $S_\textnormal{inel}(\mathbf{q},\omega)$ are shown in Fig.~\ref{fig1} (b). The inset shows a magnified segment around the $\omega\to0$ limit, which exhibits an unexpected drop for $\omega\lesssim1\,$eV. This reflects the absence of a continuum (or quasi-continuum) portion of the spectra that contributes to thermal transitions with small energy differences, as TDDFT modeling has computational limitations in this regard~\cite{supplement}.

\begin{figure}
\centering
\includegraphics[width=0.49\textwidth]{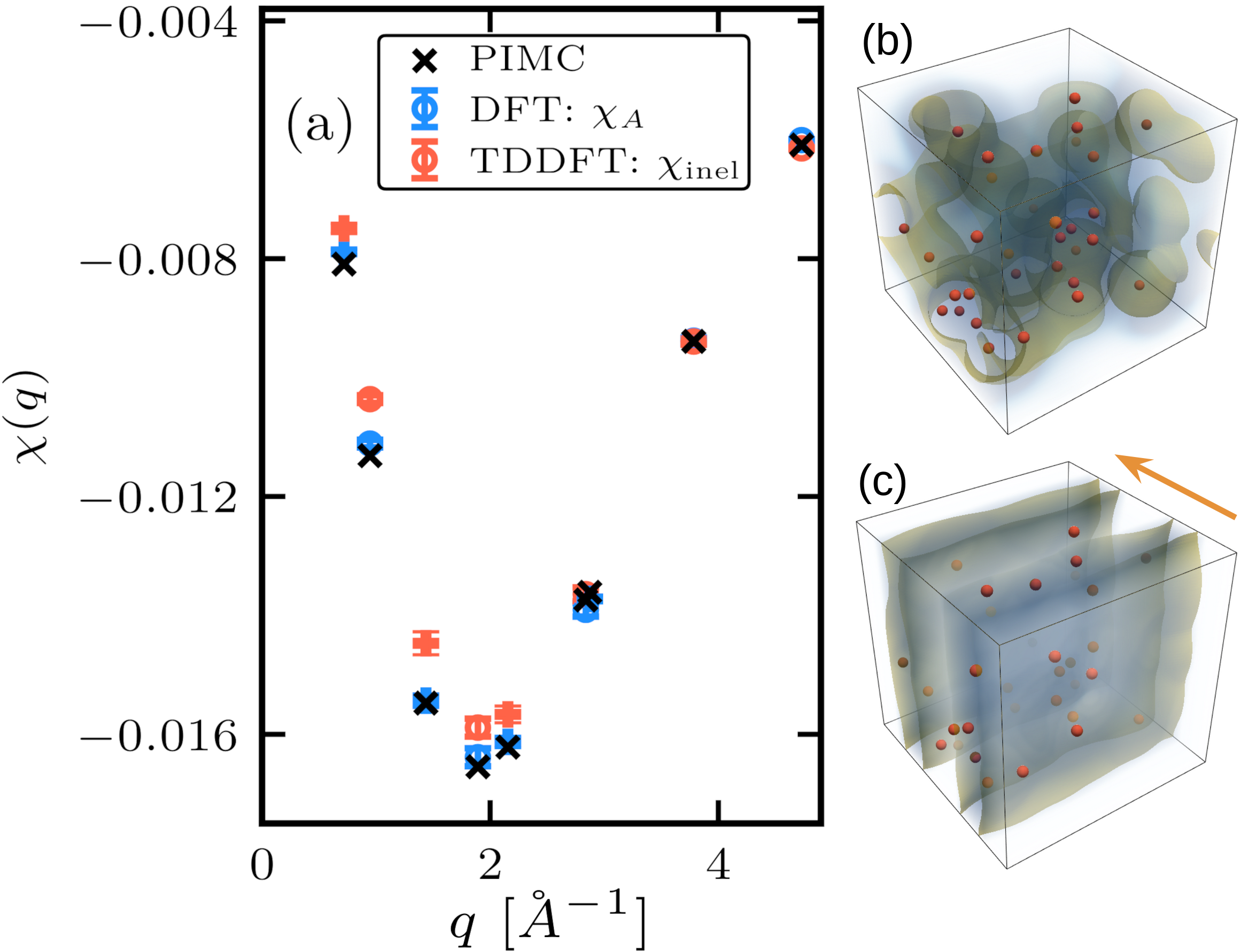}
\caption{\label{fig2} (a) Static electron density response function $\chi_\textnormal{inel}(\mathbf{q})$ [Eq.~(\ref{eq:static_chi})] of hydrogen at $\rho=0.08\,$g/cc and $T=4.8\,$eV. Black crosses: quasi-exact PIMC reference data~\cite{Dornheim_MRE_2024,Dornheim_JCP_2024}; red symbols: direct TDDFT; green symbols: direct perturbation approach, $\chi_A(\mathbf{q})$. (b) and (c): electronic isosurfaces of the perturbed and unperturbed systems; see the Supplemental Material~\cite{supplement} for details.
}
\end{figure} 

To correct for this deficiency, we consider the imaginary-time version of the fluctuation--dissipation theorem~\cite{Dornheim_MRE_2023}, which connects the the area under the ITCF with the linear density response to an external harmonic perturbation,
\begin{eqnarray}\label{eq:static_chi}
    \chi_\textnormal{inel}(\mathbf{q}) = - n_e \int_0^\beta \textnormal{d}\tau\ F_\textnormal{inel}(\mathbf{q},\tau) \, .
\end{eqnarray}
Clearly, accurate knowledge of Eq.~(\ref{eq:static_chi}) would be sufficient to correct for the constant shift observed in Fig.~\ref{fig1}.
Indeed, there is strong empirical evidence that equilibrium DFT simulations using the direct perturbation approach~\cite{Moldabekov_JCTC_2023,Moldabekov_JCTC_2022}---i.e., applying an external perturbation of wavenumber $q$ and amplitude $A$, cf.~Fig.~\ref{fig2} (b) and (c)---is capable of reproducing PIMC reference data to a remarkable degree.
In Fig.~\ref{fig2}, we show $\chi_\textnormal{inel}(\mathbf{q})$ and confirm the expected trends: the direct TDDFT results (red symbols) systematically deviate from the PIMC reference results (black crosses); in contrast, the direct perturbation results ($\chi_A(\mathbf{q}$), green symbols) circles are in excellent agreement with PIMC for all wavenumbers.
It is straightforward to estimate a corresponding correction term for both $S_{ee}(\mathbf{q})$ and the full ITCF
\begin{eqnarray}\label{eq:correction}
   \Delta F(\mathbf{q}) \equiv \Delta S_{ee}(\mathbf{q}) =  - \frac{k_\textnormal{B}T}{n_e}\left( \chi_A(\mathbf{q}) - \chi_\textnormal{inel}^\textnormal{TDDFT}(\mathbf{q})  \right)\ ,
\end{eqnarray}
which follows from the static limit of the Fourier-Matsubara series expansion~\cite{Tolias_JCP_2024}; see the Supplemental Material~\cite{supplement} for additional details. The full electron--electron static structure factor is then given by
\begin{eqnarray}\label{eq:final}
    S_{ee}(\mathbf{q}) = W_R(\mathbf{q}) + S^\textnormal{TDDFT}_\textnormal{inel}(\mathbf{q}) + \Delta S_{ee}(\mathbf{q})\, ,   
\end{eqnarray}
combining the Rayleigh weight from DFT molecular dynamics, the integral over the inelastic dynamic structure factor from TDDFT and the correction from a set of harmonically perturbed, static DFT calculations. Analogous expressions can be derived, e.g., for the full ITCF $F_{ee}(\mathbf{q},\tau)$ and for the full static linear density response function $\chi_{ee}(\mathbf{q})$, see the green curve in Fig.~\ref{fig1} (a) above.
Evaluating Eq.~(\ref{eq:final}) leads to the green circles in Fig.~\ref{fig3}. They are in very good agreement with the PIMC results for $q\lesssim3q_\textnormal{F}$, but start to disagree at around thrice the Fermi wavenumber.
This is a direct consequence of the inconsistent combination of an adiabatic (i.e., static) XC-kernel with the dynamic Kohn-Sham response function within linear-response TDDFT~\cite{Dornheim_PRL_2020_ESA}.
{The correction in  Eq.~(\ref{eq:correction}), utilizing the static density response, does not correct the inaccuracies due to the use of an adiabatic (static) XC-kernel in the TDDFT calculations.}
While these inaccuracies are relatively mild for the DSF at a given individual wavenumber, the spurious increase in $S_{ee}(\mathbf{q})$ for large $q$ accumulates for integrated properties such as the interaction energy.
The underlying inconsistency has been investigated very recently in detail for the UEG model by Dornheim \emph{et al.}~\cite{Dornheim_PRB_2024} in terms of the dynamic Matsubara density response function $\widetilde{\chi}(\mathbf{q},z_l)$, where $z_l=i2\pi k_\textnormal{B}T l$ are the discrete bosonic imaginary Matsubara frequencies.
{To eliminate the effect of the static approximation in the XC kernel, we introduce a second correction using the Fourier-Matsubara expansion of the static structure factor.}
The familiar Fourier-Matsubara expansion for $S_{ee}(\mathbf{q})$ immediately leads to the first-order correction
\begin{eqnarray}\label{eq:quantum_correction}\Delta\chi_l(\mathbf{q}) = -\frac{2k_\textnormal{B}T}{n}
\left[ 
\sum_{l=1}^\infty\left(
\widetilde{\chi}_\textnormal{RPA}(\mathbf{q},z_l) - \widetilde{\chi}_\textnormal{inel}(\mathbf{q},z_l)
\right)
\right]\ ,
\end{eqnarray}
which removes the effects of the XC-kernel from the dynamic density response except for the static limit of $\widetilde{\chi}_\textnormal{inel}(\mathbf{q},0)=\chi_\textnormal{inel}(\mathbf{q})$.
Adding $\Delta\chi_l(\mathbf{q})$ to Eq.~(\ref{eq:final})
leads to our final result for the electron static structure factor,
\begin{eqnarray}\label{eq:final_final}
    S_{ee}(\mathbf{q}) = W_R(\mathbf{q}) + S^\textnormal{TDDFT}_\textnormal{inel}(\mathbf{q}) + \Delta S_{ee}(\mathbf{q}) + \Delta\chi_l(\mathbf{q})\, .
\end{eqnarray}
The corresponding results are shown as the yellow symbols in Fig.~\ref{fig3} and are in striking agreement with the PIMC baseline over the entire range of wavenumbers.
We stress that Eq.~(\ref{eq:final_final}) does not require any external input apart from the usual electronic XC-functional, as we explain in more detail in the Supplemental Material~\cite{supplement}.



\begin{figure}
\centering
\includegraphics[width=0.47\textwidth]{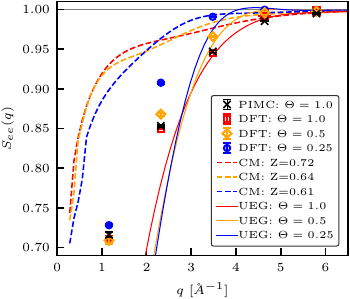}
\caption{\label{fig4} 
Electron static structure factor $S_{ee}(\mathbf{q})$ of hydrogen at $\rho=0.33\,$g/cc [$r_s=2$] and $T=12.5\,$eV [$\Theta=1$, red], $T=6.27\,$eV [$\Theta=0.5$, green] and $T=3.13\,$eV [$\Theta=0.25$, blue]. Symbols show fully corrected TDDFT data [Eq.~(\ref{eq:final_final})] and the curves have been computed based on a chemical model~\cite{bellenbaum2025estimatingionizationstatescontinuum}, see the Supplemental Material~\cite{supplement} for details. The black crosses show quasi-exact PIMC reference data~\cite{Dornheim_JCP_2024} for $\Theta=1$.
}
\end{figure}

To demonstrate the versatility of our approach, we next investigate $S_{ee}(\mathbf{q})$ at a substantially higher density of $\rho=0.33\,$g/cc over a broad range of temperatures in Fig.~\ref{fig4}. The colored symbols show our fully corrected TDDFT results, i.e., Eq.~(\ref{eq:final_final}).
At the electronic Fermi temperature of $T=12.5\,$eV, we again find excellent agreement with quasi-exact PIMC reference data~\cite{Dornheim_JCP_2024} (black crosses) without the need for any empirical parameters.
For lower temperatures, PIMC becomes unavailable due to the notorious fermion sign problem~\cite{dornheim_sign_problem,troyer}.
Overall, we again find an increase of $S_{ee}(\mathbf{q})$ of hydrogen compared to the UEG model (solid curves) for $q\lesssim3q_\textnormal{F}$ due to electronic localization around the ions. Interestingly, our TDDFT results for $S_{ee}(\mathbf{q})$ increase with decreasing temperature for intermediate wavenumber due to increasing electronic localization, whereas we find the opposite trend for the UEG.
In addition, we compare our new results to a Chihara model (dashed lines), which, however, drastically overestimates the effect of localization for all temperatures.
Taken together, Figs.~\ref{fig3} and \ref{fig4} thus also clearly highlight the importance of ab-initio simulations for the description of WDM in general, and for the interpretation of x-ray scattering experiments~\cite{Dornheim_SciRep_2024,siegfried_review,dornheim2024modelfreerayleighweightxray}, in particular.

\textbf{Discussion.} In this work, we have introduced a new framework for the computation of the electron static structure factor $S_{ee}(\mathbf{q})$ from thermal DFT simulations. This has been achieved by combining the Rayleigh weight $W_\textnormal{R}(\mathbf{q})$, the static linear density density response $\chi_\textnormal{inel}(\mathbf{q})$ from a harmonically perturbed equilibrium calculation, and the frequency dependence from the dynamic Kohn-Sham response function. In addition, we remove the spurious effect of an adiabatic XC-kernel onto $S_{ee}(\mathbf{q})$ with a dynamic Matsubara correction, Eq.~(\ref{eq:quantum_correction}).
The thus computed DFT results for $S_{ee}(\mathbf{q})$ are in excellent agreement with quasi-exact PIMC reference data, where they are available, without the need for any free empirical parameters.
In addition, we presented DFT results for hydrogen at significantly lower temperatures, where PIMC is unavailable due to the sign problem.

We are convinced that our work opens up a variety of avenues for important future research across research fields.
First and foremost, we stress that the present investigation of warm dense hydrogen is important in its own right. For example, we have unambiguously demonstrated the substantial and systematic deficiency of ubiquitous chemical models, which constitute a widely used method for the interpretation of x-ray scattering experiments with WDM~\cite{Tilo_Nature_2023,boehme2023evidence,Poole_PRR_2024,kraus_xrts}, for different densities and temperatures.

{
The results presented were computed using the PBE XC functional. Generally, the methodology discussed can be applied in conjunction with any other XC functional. Depending on the type of material and the thermodynamic conditions, it may be necessary to use higher-rank XC functionals. Therefore, the limitations of the presented method are defined by the ability of the chosen XC functional to describe both electronic and atomic structures accurately. For instance, describing dense hydrogen in regions where molecular formations are predominant is expected to be more challenging than the case considered, which involves a significant degree of ionization.}

Beyond hydrogen, we note that $S_{ee}(\mathbf{q})$ constitutes an observable in both x-ray diffraction and x-ray Thomson scattering experiments~\cite{Dornheim_SciRep_2024,siegfried_review,sheffield2010plasma}; our new DFT set-up will thus be useful to predict the outcome of future experiments and to facilitate the interpretation of existing measurements~\cite{dornheim2024modelfreerayleighweightxray,Dornheim_SciRep_2024,Dornheim_Science_2024}.
Indeed, a key advantage of DFT over the potentially more accurate but computationally more expensive PIMC approach is that it is available for a host of heavier elements and composite materials, and over a broad range of densities and temperatures. 

Finally, we highlight the new opportunities for further  methodological developments. For example, accurate DFT results for $S_{ee}(\mathbf{q})$ can be used to separately estimate the interaction energy $W$ and, in this way, to split the total energy into its potential and kinetic contributions. In addition, $S_{ee}(\mathbf{q})$ and $W$ constitute important ingredients for the construction of advanced XC-functionals based on the adiabatic connection formula~\cite{Aurora_prl_2016}.
Finally, it will be straightforward to combine the present approach with existing and upcoming dynamic XC-kernels both in the real-frequency and imaginary Matsubara frequency domain.

\begin{acknowledgements}

\noindent This work was partially supported by the Center for Advanced Systems Understanding (CASUS), financed by Germany’s Federal Ministry of Education and Research and the Saxon state government out of the State budget approved by the Saxon State Parliament. This work has received funding from the European Union's Just Transition Fund (JTF) within the project \emph{R\"ontgenlaser-Optimierung der Laserfusion} (ROLF), contract number 5086999001, co-financed by the Saxon state government out of the State budget approved by the Saxon State Parliament. This work has received funding from the European Research Council (ERC) under the European Union’s Horizon 2022 research and innovation programme (Grant agreement No. 101076233, "PREXTREME"). 
Views and opinions expressed are however those of the authors only and do not necessarily reflect those of the European Union or the European Research Council Executive Agency. Neither the European Union nor the granting authority can be held responsible for them. Computations were performed on a Bull Cluster at the Center for Information Services and High-Performance Computing (ZIH) at Technische Universit\"at Dresden and at the Norddeutscher Verbund f\"ur Hoch- und H\"ochstleistungsrechnen (HLRN) under grant mvp00024.
WM acknowledges support by the National Natural Science Foundation of China under Grant No. 12274171. XS acknowledges support by the Advanced Materials-National Science and Technology Major Project (No. 2024ZD0606900).
\end{acknowledgements}

\section*{Author Declarations}
\subsection*{Conflict of interest}
The authors have no conflicts to disclose.

\bibliography{bibliography}
\end{document}